# Embedding Lithium-ion Battery Scrapping Criterion and Degradation Model in Optimal Operation of Peak-shaving Energy Storage

Qingchun Hou, Yanghao Yu, Ershun Du,
Hongjie He, Ning Zhang, Chongqing Kang
Department of Electrical Engineering
Tsinghua University, Beijing, China

Guojing Liu,
Huan Zhu
Economic Research Institute
Jiangsu Electric Power Company, Nanjing, China

*Abstract*—Lithium-ion battery systems have been used in practical power systems for peak-shaving, demand response, and frequency regulation. However, a lithium-ion battery is degrading while cycling and would be scrapped when the capacity reduces to a certain threshold (e.g. 80%). Such scrapping criterion may not explore the maximum benefit from the battery storage. In this paper, we propose a novel scrapping criterion for peak-shaving energy storage based on battery efficiency, time-of-use price, and arbitrage benefit. A new battery life model with scrapping parameters is then derived using this criterion. Embedded with the life model, an optimal operation method for peak-shaving energy storage system is presented. The results of case study show that the operation method could maximize the benefits of peak-shaving energy storage while delaying battery degradation. Compared with the traditional 80% capacity-based scrapping criterion, our efficiency-based scrapping criterion can significantly improve the lifetime benefit of the battery.

## I. INTRODUCTION

The burden of power system peak-shaving has been sharply increasing due to the mismatch between peak load and renewable energy generation and the shortage of flexible resources [1][2][3]. To ease the burden, more energy storage systems are needed to improve power system flexibility [4][5]. Lithium-ion battery systems have been used in practical power systems for peak-shaving [6], demand response [7], frequency regulation [8][9][10], and renewable energy fluctuations suppression [11][12]. Meanwhile, lithium-ion battery degrades over time and cycles [13]. The corresponding battery life is called calendar and cycle life, respectively. The degradation is mainly caused by two factors: 1) loss of lithium-ions due to solid electrolyte interface (SEI) formation; 2) loss of electrode sites [14]. These changes increase internal resistance, decrease capacity and efficiency, and eventually shorten the battery life [15][16]. Therefore, the degradation inevitably affects the optimal operation and lifetime benefit of lithium-ion battery energy storage, especially with increasing energy storage penetration in power system. It's in urgent need to model lithium-ion battery degradation, determine the battery end of life, and consider battery degradation cost in grid-connected energy storage operation.

Researchers have developed several models to uncover the battery degradation mechanism. Xu et al. classified the lithium-ion battery life model into theoretical models and empirical models [17]. The theoretical models focus on the loss of active materials and explanation of degradation mechanism, while the empirical models are easier to embed in operation and planning research. They further proposed a new empirical stress model for cycle loss of three types of lithium-ion batteries. Wang et al. established a capacity loss model for graphite-LiFePO4 battery [14]. The capacity loss is a power-law function of charge throughput and an Arrhenius function of temperature. Redondo-Iglesias et al. found that there is high dependence between efficiency decrease and capacity fade in calendar life of lithium-ion battery for electric vehicles [18]. They proposed a battery degradation model based on the correlation between capacity fade and efficiency decrease. However, degradation of grid-connected lithium-ion battery has been largely ignored because it is difficult to embed the complex and non-linear battery life model in optimization [11][19][20][21][22]. In addition, Mishra et al. conclude that the lithium-ion battery life varies significantly for different energy storage application scenarios such as time-of-use energy management, solar self-consumption, and power backup [23].

Recently, many researchers try to consider the battery degradation in optimization and integrate battery life model into different power system applications [24]–[26]. For instance, Maheshwari et al. developed a non-linear lithium-ion battery degradation model from experimental data. They embedded the model into energy storage optimal operation by making it compatible with a MILP formulation [26]. Li et al. integrated the lithium-ion battery degradation such as capacity fade into power flow model. The model aims to make the renewable energy dispatchable using energy storage system [27]. Liu et al. introduced the capacity loss model w.r.t time, depth of discharge (DOD), and charge throughout [28]. They further simplified the model to incorporate it in capacity planning of lithium-ion battery considering PV generation. Shi et al. assumed that the lithium-ion battery has a constant marginal degradation cost [6]. The degradation cost is included in the objective of battery operation model for peak shaving and frequency regulation. They concluded that the benefit of joint optimization is higher than the sum of individual benefits from peak shaving and frequency regulation. He et al. used power function to model the relationship between maximum battery cycle number and DOD, and then derived the loss of battery life [29]. They applied the model into battery operation for frequency regulation on electricity market. Tran, et al. proposed a battery life model for micro-grid application based on lifetime energy throughput. The results show the model could improve both the energy storage efficiency and battery life [12]. Shi, et al. applied Rainflow algorithm to identify battery cycle number and embedded the cycle-based cost into optimization model. They proved the model is convex and can extend the lifetime of battery [30].

This work was supported by Scientific & technical project of State Grid (No. 5102-201918309A-0-0-00).
Corresponding author: Ning Zhang (ningzhang@tsinghua.edu.cn) and Chongqing Kang (cqkang@tsinghua.edu.cn).

Previous researches about embedding battery degradation in optimization mainly focus on capacity decrease. In contrast, the efficiency decrease, especially for cycle life, is less studied [18]. For example, Ahmadi et al. applied used lithium-ion battery pack for power system application and concluded that the efficiency is important for re-used lithium-ion battery. Due to the lack of efficiency data, they assumed that efficiency decrease has the same trend as capacity fade [16]. In addition, the scrapping criterion for grid-connected energy storage system is seldom discussed and most researches use a capacity-based criterion, such as 80% capacity to determine the end of life (EOL) of lithium-ion battery [15]. This criterion may be suitable for electrical vehicles because high power is required for extreme traffic scenarios. However, it is hardly effective for grid-connected battery. After reaching 80% capacity, the grid-connected battery can still benefit from electricity market for peak-shaving and frequency regulation. The new scraping criterion and degradation model are necessary to explore maximum benefit from grid-connected lithium-ion battery energy storage system.

To bridge the gap, this paper proposes a novel efficiency-based lithium-ion battery scrapping criterion for peak-shaving energy storage system to explore maximum lifetime benefit from the battery. This criterion can be used for both new and re-used battery in power system peak shaving applications. We also present a battery life model using the proposed criterion. In the model, maximum cycle number of the battery is derived as a function of DOD and scrapping parameters to make the model easy to embed in optimization. Furthermore, we proposed an optimal operation model for peak-shaving energy storage system considering the battery degradation. This model can maximize the benefit of peak-shaving while minimizing the cost of battery degradation. Therefore, the contributions of this paper are as follows:

(1) We propose an efficiency-based lithium-ion battery scraping criterion for peak-shaving energy storage to explore maximum lifetime benefit from lithium-ion battery;

(2) We propose a new battery degradation model with scraping parameters;

(3) We embed the battery degradation model in energy storage operation optimization to maximize the battery lifetime benefit. A lifetime benefit comparison between efficiency-based and capacity-based scraping criterions is conducted using Jiangsu province data in China.

The remainder of the paper is organized as follows. Section II presents the methodology framework, the new scrapping criterion, and battery life model. Section III introduces the energy storage operation optimization model. Section IV presents a case study using Jiangsu province data in China for validation and compares the benefit between efficiency-based and capacity-based degradation model. Section V concludes the paper and describes future work.

## II. METHODOLOGY

### A. Framework

There are three main issues related to lithium-ion battery optimal operation: 1) how to quantify the EOL of lithium-ion battery? 2) How to model the battery life and embed it in operation optimization with different scrapping criteria? 3) How to make the operation optimization model easy to solve? To this end, Fig. 1 shows the framework of our three-stage method. First, we propose a new scrapping criterion based on battery efficiency and time-of-use prices. Second, we derived a battery life model using this criterion. The model describes the relationship among maximum cycle number, DOD, and the scrapping parameter. Third, the maximum cycle number is expressed as the loss rate of the battery and then embedded in the operation optimization model. The objective of the optimization model is to maximize the benefit of peak-shaving and minimize the cost of battery degradation. The constraints of energy storage, renewable energy, and power balance are also taken into consideration. To make the optimization problem easy to solve, degradation cost is approximated by multiplying the cycle loss rate and investment cost of energy storage.

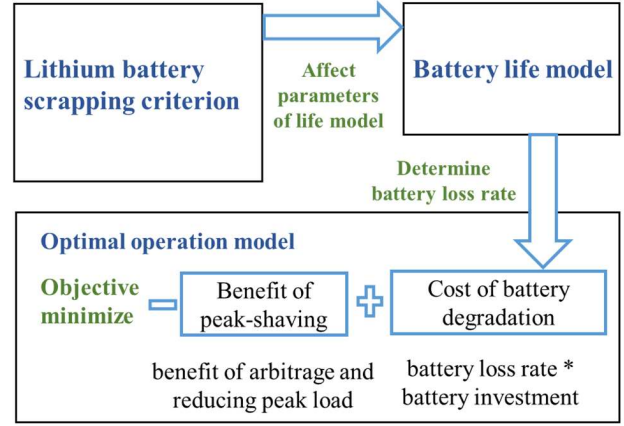

Figure 1. The framework of the method

### B. Efficiency-based Scrapping Criterion

Instead of 80% of rated capacity, our lithium-ion battery scrapping criterion for peak-shaving energy storage is based on battery efficiency, time-of-use prices, and arbitrage benefit. The core idea is that the battery should be scrapped when the arbitrage benefit of peak-shaving battery energy storage cannot balance the battery operation and maintenance (O&M) cost:

$$\pi^p E \eta^{dis} - \pi^v E / \eta^{cha} < \pi^{cs}(E\eta^{dis} + E / \eta^{cha})$$
$$\Rightarrow \eta^{dis}\eta^{cha} < \frac{\pi^v + \pi^{cs}}{\pi^p - \pi^{cs}} \quad (1)$$

where $E$ is electricity charged into the battery in the valley load time of power system. $\pi^p$ and $\pi^v$ are electricity prices in the peak load time and the valley load time, respectively. $\eta^{dis}$ and $\eta^{cha}$ are the discharge and charge efficiency of the battery, respectively. $\pi^{cs}$ is the average O&M cost for unit charge or discharge. It should be noted that the efficiency of energy storage here is not coulombic efficiency but energy efficiency.

When incorporating the scrapping criterion into battery life model, the most difficult part is to model the efficiency decrease of lithium-ion battery, which is seldom studied in the cycle life [18]. Thus, we try to establish a relationship among average energy efficiency, capacity, and resistance. Fig. 2 illustrates the equivalent circuit of a lithium-ion battery and how it is connected to the grid by inverter. We assume that lithium-ion battery charges and discharges with $n^{cha}$ and $n^{dis}$ C current, respectively. $V^{cha}$ and $V^{dis}$ are the average voltage of charge and discharge, respectively. Then, the scrapping criterion can be expressed in terms of internal resistance $R$, and capacity $C_L$ (with the unit of Ah) as follows:

$$\eta^{total} = \eta^{inv}\eta^{dis}\eta^{cha} = \eta^{inv}\frac{P^{dis}}{P^{cha}}$$
$$= \eta^{inv}\frac{n^{dis}C_L V^{dis} - (n^{dis}C_L)^2 R}{n^{cha}C_L V^{cha} + (n^{cha}C_L)^2 R} \quad (2)$$
$$= \eta^{inv}\frac{n^{dis}}{n^{cha}}\cdot\frac{V^{dis} - n^{dis}C_L R}{V^{cha} + n^{cha}C_L R} < \frac{\pi^v + \pi^{cs}}{\pi^p - \pi^{cs}}$$

where $\eta^{inv}$ is the average overall efficiency of the inverter. $\eta^{total}$ is the total efficiency of inverter and battery. $n^{dis}C_L V^{dis} = n^{cha}C_L V^{cha}$ should hold in (2) to ensure the balance between output power and input power.

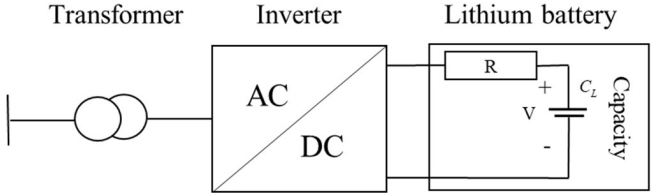

Figure 2. Illustration of lithium-ion battery connected to the grid.

Simplifying equation (2), the lithium-ion battery reaches EOL when:

$$C_L R > \frac{V^{dis} - y V^{cha}}{n^{dis} + y n^{cha}}$$
$$y = \frac{(\pi^v + \pi^{cs})n^{cha}}{(\pi^p - \pi^{cs})\eta^{inv} n^{dis}} \quad (3)$$

Equation (3) transforms the original efficiency-based criterion to a capacity and resistance-based criterion.

C. Lithium-ion Battery Life Model

The cycle degradation of normalized capacity $C_L^*$ and normalized resistance $R^*$ is a function of DOD and charge throughput [13]:

$$C_L^* = 1 - (\beta_0 + \beta_1 d)\cdot\sqrt{Q} \quad (4)$$
$$R^* = 1 + (\alpha_0 + \alpha_1 d)\cdot Q \quad (5)$$

where $d$ is the DOD. $Q$ is the charge throughput. $\beta_0$, $\beta_1$, $\alpha_0$, $\alpha_1$ are calculated as:

$$\beta_0(V) = 7.348\cdot 10^{-3}\cdot(V - 3.667)^2 + 7.600\cdot 10^{-4}$$
$$\beta_1 = 4.081\cdot 10^{-3}$$
$$\alpha_0(V) = 2.153\cdot 10^{-4}\cdot(V - 3.725)^2 - 1.521\cdot 10^{-5} \quad (6)$$
$$\alpha_1 = 2.798\cdot 10^{-4}$$

where $V$ is the average voltage of charge and discharge.

To derive the battery life using maximum cycle number, we approximately express $Q$ by cycle number and DOD:

$$Q \approx N\cdot 2d\cdot C^{st} \quad (7)$$

where $N$ is the cycle number, $C^{st}$ is the battery initial capacity.

1) Life model with capacity-based scrapping criterion

Substituting (7) into (4), the function between cycle number $N$ and capacity degradation is:

$$N = \frac{(1 - C_L^*)^2}{2C^{st}d(\beta_0 + \beta_1 d)^2} \quad (8)$$

Thus, the maximum cycle number $N^{end}$ with capacity degradation criterion $C^{end}$ is:

$$N^{end} = \frac{(1 - C^{end})^2}{2C^{st}d(\beta_0 + \beta_1 d)^2} \quad (9)$$

Equation (9) establishes the relationship between life model and capacity-based scraping criterion.

2) Life model with efficiency-based scrapping criterion

Multiplying (4) by (5), the function between $C_L^* R^*$ and $Q$ is:

$$C_L^* R^* = 1 - (\beta_0(V) + \beta_1 d)\cdot\sqrt{Q} + (\alpha_0(V) + \alpha_1 d)\cdot Q \quad (10)$$
$$- (\beta_0(V) + \beta_1 d)\cdot(\alpha_0(V) + \alpha_1 d)\cdot Q\sqrt{Q}$$

Equation (10) shows that $C_L^* R^*$ and $\sqrt{Q}$ have a cubic relationship. Thus, there is an analytical-form solution for $\sqrt{Q}$ expressed by $C_L^* R^*$, denoted as:

$$\sqrt{Q} = f^{cubic}(R^* C_L^*, d) \quad (11)$$

where $f^{cubic}$ is a real and positive solution of cubic function (10). In this paper, this solution function $f^{cubic}$ is calculated by Mathematica.

Substituting (7) into (10), the maximum cycle number $N^{end}$ with new scrapping criterion $C^{end} R^{end}$ is:

$$N^{end} = \frac{\left[f^{cubic}(C^{end}R^{end}, d)\right]^2}{2d\cdot C^{st}} \quad (12)$$

where the $C^{end} R^{end}$ is determined by (3).

Equation (12) establishes the relationship between life model and efficiency-based scraping criterion.

D. Degradation Cost of Lithium-ion Battery

Because the relationship shown in equations (9) and (12) is complex, it is difficult to embed life model into battery operation optimization. To this end, we express battery degradation as the loss rate of battery life [29]. The loss rate $f^{cycle}(n,d)$ after $n$ cycles with DOD $d$ can be derived as:

$$f^{cycle}(n,d) = \frac{n}{N^{end}(d)} \quad (13)$$

Battery reaches end of life when the accumulated loss rate is "1". Therefore, the battery cycle degradation cost $J^{loss}(n,d)$ can be expressed as the product of loss rate and total investment cost:

$$J^{cycle}(n,d) = f^{cycle}(n,d)\cdot\pi^{inv} \quad (14)$$

where $\pi^{inv}$ is the investment cost of the lithium-ion battery.

Similarly, the calendar life loss rate $f^{cal}$ and the corresponding degradation cost $J^{cal}$ in an operation day can be calculated as follows:

$$f^{cal} = \frac{1}{T^{end}} \qquad (15)$$
$$J^{cal} = f^{cal} \cdot \pi^{inv}$$

where the $T^{end}$ is constant calendar life. The calculation method can be found in [13].

### III. OPERATION OPTIMIZATION MODEL

#### A. Objective function

Our aim is to maximize the peak-shaving benefit of energy storage system with PV and load while delaying the battery degradation. Therefore, the objective of our operation optimization model, shown in (16), is to minimize the sum of four terms: 1) the cost of buying electricity from grid, 2) the cost of peak capacity, 3) the O&M cost of lithium-ion battery, and 4) the degradation cost of lithium-ion battery. The benefits of peak-shaving energy storage come from reducing the cost of the first two terms.

$$J^{day} = \min \sum_{t=1}^{M} \pi_t^g P_t^g + \pi^c L^{pk} + \sum_{i=1}^{K}\left(\sum_{t=1}^{M} \pi^{cs}(P_{i,t}^{dis} + P_{i,t}^{cha})\right) + \sum_{i=1}^{K}\left(\sum_{t=1}^{M} J_{i,t}^{cycle}(0.5, d_{i,t}) + J_i^{cal}\right) \qquad (16)$$

where $J^{day}$ is the daily total cost. $M$, $K$ are the numbers of time intervals and batteries, respectively. $\pi_t^g$, $\pi^c$ are the time-of-use electricity price at time $t$ and the peak capacity price, respectively. $P_t^g$, $L^{pk}$ are the power from the grid at time $t$ and peak capacity during a day, respectively. $P_{i,t}^{dis}$, $P_{i,t}^{cha}$, $d_{i,t}$, $J_{i,t}^{cycle}$, $J_i^{cal}$ are the discharge power, charge power, DOD, cycle degradation cost, and calendar degradation cost of $i$th battery at time $t$, respectively.

#### B. Constraints

The operation optimization model considers power balance constraint, energy storage constraints, and renewable energy constraints.

The power balance constraint is the power from the grid $P_t^g$, PV generation $P_t^r$, and energy storage output $P_{i,t}^{dis} - P_{i,t}^{cha}$ should meet the load $L_t$ at time $t$.

$$P_t^g + P_t^r + \sum_i^K (P_{i,t}^{dis} - P_{i,t}^{cha}) = L_t \quad \forall t = 1 \cdot \cdot M \qquad (17)$$

The energy storage has constraints of maximum SOC and power output.

$$\begin{aligned} 0 &\leq S_{i,t} \leq C_i^b \\ 0 &\leq P_{i,t}^{dis} \leq \overline{p}_i^{dis} \\ 0 &\leq P_{i,t}^{cha} \leq \overline{p}_i^{cha} \\ \forall t &= 1 \cdot \cdot M, i = 1 \cdot \cdot K \end{aligned} \qquad (18)$$

where $C_i^b$, $\overline{p}_i^{dis}$, $\overline{p}_i^{cha}$ are energy capacity, maximum discharge and charge power of $i$th battery, respectively. $S_{i,t}$ is the SOC of $i$th battery at time $t$.

The relationship between SOC and energy storage net output is as follows:

$$\begin{aligned} P_{i,t}^b &= P_{i,t}^{dis}/\eta_{i,t}^{dis} - P_{i,t}^{cha}\eta_{i,t}^{cha} \\ S_{i,t} - S_{i,t-1} &= -P_{i,t}^b \\ \forall t &= 1 \cdot \cdot M, i = 1 \cdot \cdot K \end{aligned} \qquad (19)$$

where $P_{i,t}^b, \eta_{i,t}^{dis}, \eta_{i,t}^{cha}$ are net output, discharge and charge efficiency of $i$th battery at time $t$, respectively.

The DOD is calculated as follows:

$$d_{i,t}^b = \left|P_{i,t}^b / C_i^b\right| \quad \forall t = 1 \cdot \cdot M, i = 1 \cdot \cdot K \qquad (20)$$

The output of renewable energy should be less than the day-ahead prediction $\overline{p}_t^r$.

$$0 \leq p_t^r \leq \overline{p}_t^r \quad \forall t = 1 \cdot \cdot M \qquad (21)$$

The peak capacity is the maximum load from the grid during a day:

$$L^{pk} = \max(P_t^g) \qquad (22)$$

For real-world power systems in China, selling electricity to the grid from the battery-PV system is not allowed currently.

$$P_t^g \geq 0 \quad \forall t = 1 \cdot \cdot M \qquad (23)$$

#### C. Solving Method

We linearize equation (13) using piece-linearized technique and solve the operation optimization model with Gurobi. The solving method is listed in Algorithm 1.

| **Algorithm 1:** Operation optimization of peak-shaving storage | |
|---|---|
| **input** | Battery parameters (initial capacity and efficiency, average discharge and charge voltage, rated charge and discharge current), load, PV prediction, time-of-use prices |
| **Step 1** | Determine the EOL using (2)-(3) or 80% capacity |
| **Step 2** | Determine parameters of life model using (9) or (12) |
| **Step 3** | Derive the degradation cost using (13)-(15) and piece-linearize degradation cost function |
| **Step 4** | Solve the linear operation optimization model (16)-(23) with Gurobi |
| **Output** | Battery output, PV generation, Cost of buying electricity, peak capacity, O&M, and degradation |

#### D. Lifetime Benefit Estimation

According to (13)-(15), a battery reaches EOL when the sum of loss rate is "1". Therefore, the lifetime $T^{total}$ (the unit is day) can be expressed as:

$$T^{total} = \frac{1}{\sum_{t=1}^{M} J_t^{cycle}(0.5, d_t) + J^{cal}} \qquad (24)$$

Then, the total benefit $J^{total}$ of the battery can be estimated as the product of typical daily benefit and lifetime:

$$J^{total} = \left(J^0 - J^{day}\right) \cdot T^{total} \quad (25)$$

where $J^0$ is the daily operation cost without battery.

It should be noted that (24) is an estimation of battery lifetime when using average capacity and efficiency in daily operation. Accurate lifetime can be obtained by simulating battery operation iteratively until the battery reaches the EOL, which computation burden is much heavier.

## IV. CASE STUDY

We consider a 4 MW / 4 MWh lithium-ion energy storage system with 12 MW PV in Jiangsu Province, China. In the rest of the section, we will validate the proposed methods with analysis and comparison of case study results.

### A. Data Description

In this case study, real-world data such as time-of-use prices, PV generation, and load profile are from Jiangsu Province of China. The load profile and typical PV prediction of a typical day are shown in Fig. 3. The load mainly has two peaks during 9:00-17:00 and 20:00-22:00. The PV generation peaks at 13:00. Noted that the uncertainty of PV prediction can be considered using stochastic programing with typical generation scenarios [6]. The time-of-use prices and peak capacity price for industrial load are listed in Table I. We use 18650 lithium-ion battery pack that is rated for 300-500 full cycles. The parameters for the battery EOL calculation are shown in Table II. We assume that all batteries in the pack have the same degradation trend. Thus, the average voltage and current in Table II are parameters of a single battery. The investment cost of lithium-ion battery is 176 \$/kWh [31].

To study how the degradation and scrapping criterion affect the operation and peaking-shaving benefits of battery, four scenarios are studied. Scenario 1: without energy storage; Scenario 2: with energy storage but ignoring degradation; Scenario 3: with energy storage and 80% capacity-based scrapping criterion, and Scenario 4: with energy storage and efficiency-based scrapping criterion. Calculated by Equation (2) with parameters in Table I, the overall scraping efficiency is 61.6%. The overall efficiency is defined as the product of charge and discharge efficiency.

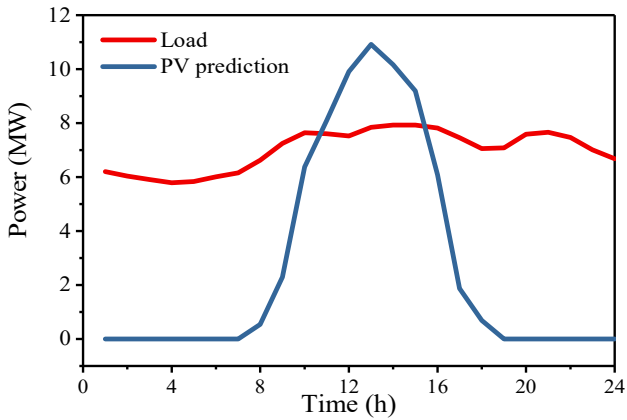

Figure 3. Load and PV prediction of a typical day

TABLE I THE TIME-OF-USE PRICES AND PEAK CAPACITY PRICE FOR INDUSTRIAL AND COMMERCIAL LOAD IN JIANGSU

|  | Peak energy /kWh | Normal energy /kWh | Valley energy /kWh | Monthly Peak capacity /kW |
|---|---|---|---|---|
| Price /\$ | 0.153 | 0.092 | 0.05 | 10 |
| Time | 8:00-12:00 17:00-21:00 | 12:00-17:00 21:00-24:00 | 0:00-8:00 | – |

TABLE II THE PARAMETERS OF CALCULATING BATTERY EOL

| The charge and discharge efficiency of battery | The efficiency of inverter | Average voltage /V | Charge and discharge current / C | Average O&M cost /(\$/ kWh) |
|---|---|---|---|---|
| 89% | 90% | 3.7 | 1 | 0.017 |

### B. Results of Battery Life Model

Fig. 4 and Fig. 5 show how capacity and efficiency degrade over cycle number with a certain DOD, respectively. Both capacity and efficiency of lithium-ion battery decrease when cycle number increases, but the decreasing rate for efficiency is much slower. For example, after 2000 full DOD cycles, the capacity drops from 100% to 50% while the efficiency declines from about 80% to 60%. Besides, smaller DOD can significantly slow down the decrease in capacity and efficiency.

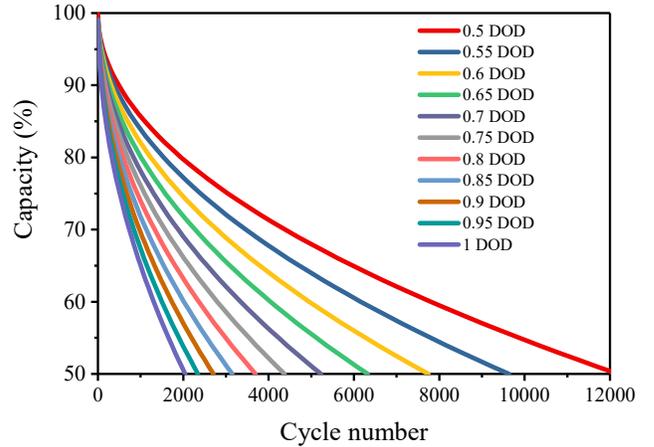

Figure 4. The degradation of battery capacity with increasing cycle number

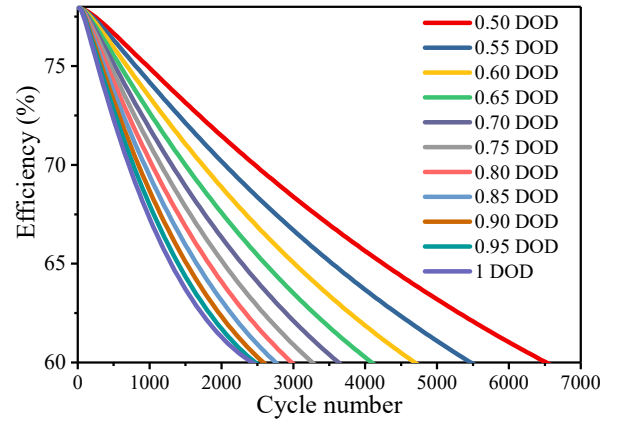

Figure 5. The degradation of battery overall efficiency with increasing cycle number

We further show the relationship between maximum cycle number and DOD using either capacity-based (Fig. 6) or efficiency-based scrapping criterion (Fig. 7). Under 50% capacity-based or 60% efficiency-based scrapping criterion, reducing DOD from 1 to 0.5 can extend the battery lifetime approximately four times and two times, respectively. The maximum cycle number is significantly affected by the scrapping criterion. For example, both reducing the scrapping capacity from 80% to 50% and reducing scrapping efficiency from 75% to 60% can extend the lifetime of battery nearly five times. As shown in Fig. 6 and Fig. 7, the relationship between the maximum cycle number and DOD is convex. This feature makes it easy to linearize the maximum cycle number function and embed it in the optimization model.

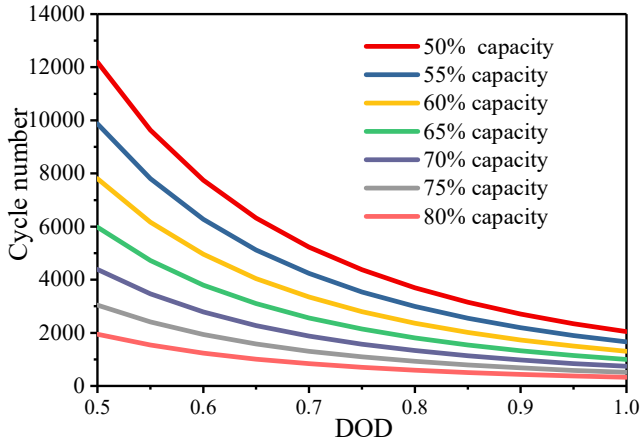

Figure 6. The relationship between maximum cycle number and depth of discharge given capacity scrapping criterion

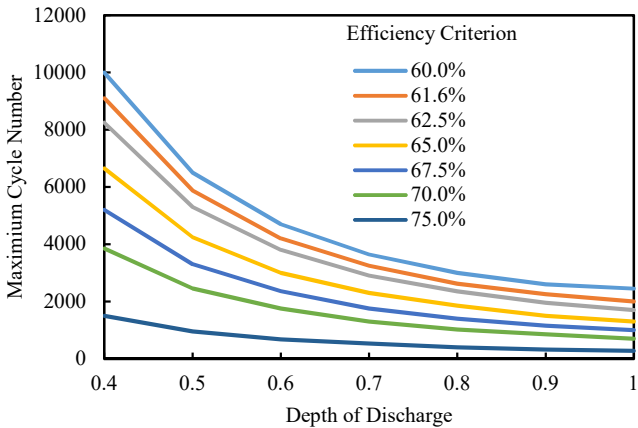

Figure 7. The relationship between maximum cycle number and depth of discharge given efficiency scrapping criterion

## C. Results of Optimal Operation

Fig. 8 shows the battery output, load, PV generation, power from the grid, PV prediction from four scenarios in a typical day. In Scenario 1, the local load is mainly served by power from the grid and PV generation. Some PV generation will be curtailed at noon due to overgeneration. In Scenario 2, the battery mainly charges during valley load time (3:00-7:00) and PV overgeneration time (10:00-16:00), and discharges during peak load time (9:00-10:00 and 19:00-23:00). In Scenario 3 and 4, the charge and discharge time are similar to Scenario 2. However, considering the trade-off between peak-shaving benefit and battery degradation, the batteries in Scenario 3 and 4 tend to charge and discharge with much smaller DOD and less frequently to extend the lifetime of battery. Fig. 9 further shows the DOD of Scenarios 2, 3, 4. The DOD in Scenario 2 is predominantly between 0.3-0.9, while DOD in Scenario 3 and 4 are mainly between 0.1-0.6 and 0.1-0.5, respectively. The number of high DOD cycles is also smaller in Scenario 3 and 4.

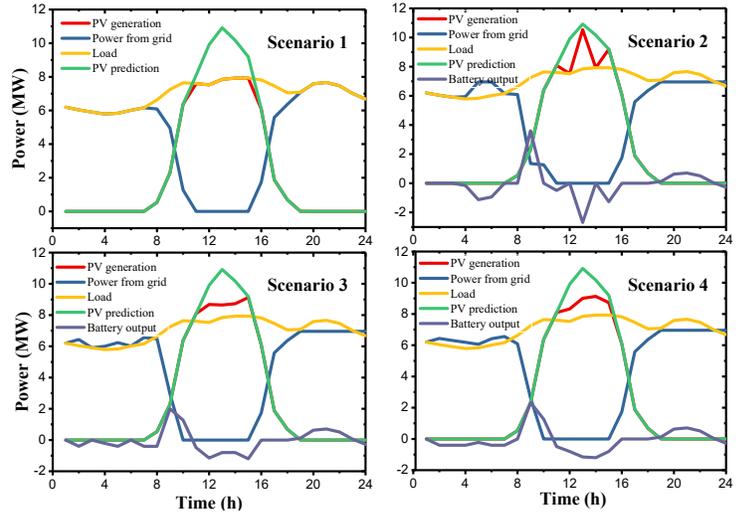

Figure 8. The battery output, load, PV generation, power from grid, PV prediction of four scenarios.

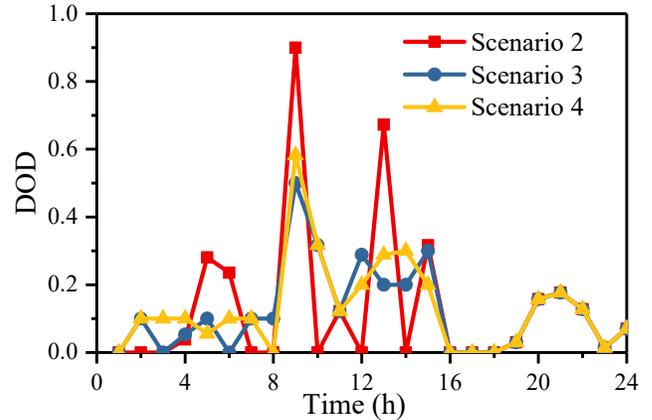

Figure 9. The depth of discharge of scenarios 2, 3, and 4.

Table III shows the daily and lifetime benefits of battery in four scenarios.

1. Comparing Scenario 2 with Scenario 1, the daily cost decreases from 12918 $ to 12150 $ by 768 $. The benefit in Scenario 2 mainly comes from the reduction of energy cost (689 $) and peak capacity cost (236 $) due to peak-shaving energy storage. The peak capacity is reduced by 0.7 MW from 7.66 MW to 6.96 MW. However, when the investment cost is taken into consideration, the lifetime benefit reduces to -315161 $. This means that the benefits from arbitrage and peak-shaving cannot even cover the investment. The main reason is that battery has a short lifetime (506 days) due to high DOD cycles and thus fast degradation speed.

2. Comparing Scenario 3 with Scenario 2, the daily benefit reduces from 768 $ to 580 $ due to lower DOD cycles, higher energy cost, and the consideration of degradation cost in daily operation. However, the lifetime benefit increases from -315161 $ to 690380 $ because the lifetime increases by nearly 100% to 1052 days. This means that considering the proposed degradation model in battery optimal operation can extend battery lifetime and increase lifetime benefit.

3. Comparing Scenario 4 with Scenario 3, the daily benefit slightly increases from 580 $ to 670 $, while the lifetime benefit sharply increases by nearly 100%, from 609380 $ to 1199935 $. The daily benefit increase is mainly due to the decreasing degradation cost from 167 $ to 98 $. The lifetime benefit increase is attributed to both higher daily benefit and longer lifetime (from 1052 days to 1792 days). This result indicates that the efficiency criterion can extend battery lifetime as well as raise daily operation benefit.

TABLE III THE DAILY AND LIFETIME BENEFIT OF BATTERY IN FOUR SCENARIOS

|  | Scenario 1 | Scenario 2 | Scenario 3 | Scenario 4 |
|---|---|---|---|---|
| Daily toal cost/$ | 12918 | 12150 | 12338 | 12248 |
| Daily energy cost/$ | 10363 | 9674 | 9704 | 9674 |
| Daily O&M cost/$ | 0 | 157 | 148 | 157 |
| Daily degradation cost/$ | 0 | 0 | 167 | 98 |
| Daily peak load cost/$ | 2555 | 2319 | 2319 | 2319 |
| Daily benefit of battery/$ | - | 768 | 580 | 670 |
| Lifetime estimation/day | - | 506 | 1052 | 1792 |
| Lifetime benefit of battery/$ | - | -315161 | 609380 | 1199935 |
| Peak capacity/MW | 7.66 | 6.96 | 6.96 | 6.96 |

## V. CONCLUSION AND FUTURE WORK

This paper proposes a novel lithium-ion battery scrapping criterion for peak-shaving energy storage based on energy efficiency, time-of-use prices, and arbitrage benefit of energy storage. This criterion can be used for both new and re-used battery in power system applications to determine EOL. The maximum cycle number is derived as a function of DOD and scrapping parameters, which makes the life model easy to embed in battery operation optimization model. The results of the case study validate the proposed method and show that battery life is significantly affected by DOD and scrapping criterion. Embedding the battery degradation model into operation optimization model can maximize the lifetime benefit of the lithium-ion energy storage system, and delay the battery degradation by fewer high DOD cycles. Compared to the 80% capacity-based criterion, a suitable efficiency-based scrapping criterion can increase the battery lifetime benefit by 100% with extended battery lifetime and improved daily operation benefit.

There are some limitations of the proposed scrapping criterion and life model: 1) this criterion is more suitable for power systems with stable time-of-use prices. For real-world power systems, especially in China, the price is controlled by the government and quite stable. If it is not the case, the scrapping criterion and life model have to be changed according to the policy. In such a scenario, switching to the proposed capacity-based life model is also a good option. 2) The proposed optimal operation model cannot be applied in power system frequency regulation. The cycle number in this paper is approximately calculated because time granularity for peak-shaving is hourly based. This may not hold for frequency regulation due to frequent cycles in a short time interval. Future work aims at using the Rainflow algorithm to count cycle number and considering frequency regulation benefit in the battery operation with the proposed life model.